\documentclass{aa}
\usepackage{graphicx}
\usepackage{txfonts}

\begin{document}

\title{Optical spectroscopy of microquasar candidates \\ at low galactic latitudes
} 

\author{J. Mart\'{\i}\inst{1}
\and J.~M. Paredes\inst{2,}\thanks{CER on Astrophysics, Particle Physics
and Cosmology. Universitat de Barcelona}
\and J.~S. Bloom\inst{3,4}
\and J. Casares\inst{5}
\and M. Rib\'o\inst{6} 
\and E.~E. Falco\inst{7} 
}

\offprints{J. Mart\'{\i}, \\ \email{jmarti@ujaen.es}}

\institute{
Departamento de F\'{\i}sica, Escuela Polit\'ecnica Superior, Universidad de Ja\'en, Virgen de la Cabeza 2, 23071 Ja\'en, Spain\\
\email{jmarti@ujaen.es}
\and Departament d'Astronomia i Meteorologia, Universitat de Barcelona, Av. Diagonal 647, 08028 Barcelona, Spain\\
\email{jmparedes@ub.es}
\and Harvard Society of Fellows, 78 Mount Auburn Street, Cambridge, MA 02138, USA
\and Harvard-Smithsonian Center for Astrophysics, MC 20, 60 Garden Street, Cambridge, MA 02138, USA\\
\email{jbloom@cfa.harvard.edu}
\and Instituto de Astrof\'{\i}sica de Canarias, Calle V\'{\i}a L\'actea s/n, 38200 La Laguna, Tenerife, Canary Islands, Spain\\
\email{jcv@ll.iac.es}
\and Service d'Astrophysique, CEA Saclay, B\^at. 709, L'Orme des
Merisiers, 91191 Gif-sur-Yvette, Cedex, France\\
\email{mribo@discovery.saclay.cea.fr}
\and Smithsonian Institution, F. L. Whipple Observatory, 
P.O. Box 97, 670 Mount Hopkins, Amado, AZ 85645, USA\\
\email{falco@cfa.harvard.edu}
}

\date{Received 2 June 2003 / Accepted 22 September 2003}


\abstract{ 
We report optical spectroscopic observations of a sample of 6 low-galactic
latitude microquasar candidates selected by cross-identification of X-ray and
radio point source catalogs for $|b^{II}| \leq 5\degr$. Two objects resulted
to be of clear extragalactic origin, as an obvious cosmologic redshift has
been measured from their emission lines. For the rest, none exhibits a clear
stellar-like spectrum as would be expected for genuine Galactic microquasars.
Their featureless spectra are consistent with being extragalactic in origin
although two of them could be also highly reddened stars. The apparent
non-confirmation of our candidates suggests that the population of persistent
microquasar systems in the Galaxy is more rare than previously believed. If
none of them is galactic, the upper limit to the space density of new
\object{Cygnus~X-3}-like microquasars within 15~kpc would be $\la 1.1 \times
10^{-12}$~pc$^{-3}$. A similar upper limit for new \object{LS~5039}-like
systems within 4~kpc is estimated to be $\la 5.6 \times 10^{-11}$~pc$^{-3}$. 

}

\maketitle

\section{Introduction} \label{introduction}

Microquasars are X-ray binary systems with relativistic jets that behave as
small scale replicas of quasars. The interest in such sources for high energy
astrophysics has been reviewed by several authors (e.g., Mirabel \&
Rodr\'{\i}guez \cite{mirabel99}). The reader is also referred to Castro-Tirado
et~al. (\cite{castro01}) and Durouchoux et~al. (\cite{durouchoux02}) for
updated accounts of the recent findings. Microquasars provide excellent
laboratories for the study of the physics of accretion/ejection phenomena and
strong gravity, on time-scales more readily accessible to observers. 

The number of confirmed microquasars currently known in the Galaxy is still
relatively small: as of June 2002, only 14 objects were confirmed members with
their jets being resolved (Rib\'o \cite{ribo02}). Their population includes
systems as famous as \object{Scorpius~X-1} (the first extrasolar X-ray source
discovered) and \object{Cygnus~X-1} (the first dynamic black hole candidate).
The production of relativistic jets may be either persistent (e.g.
\object{Cygnus~X-3}, \object{LS~5039}, etc.) or transient
(\object{GRS~1915+105}, \object{V4641~Sgr}, etc.). The relativistic ejecta
have been observed to move at apparent superluminal speeds in a few systems.
Moreover, both high and low-mass X-ray binaries have been found to behave as
microquasars. In this context, the rather heterogeneous properties of
microquasars as a group make it difficult to derive statistically robust
results. In this respect, the discovery of new sources is required to drive a
deeper understanding of the phenomema.

We searched for new microquasar systems in the Galaxy by cross-identification
techniques between X-ray and radio catalogs, namely the ROSAT all sky Bright
Source Catalog (RBSC, Voges et~al. \cite{voges99}) and the NRAO VLA Sky Survey
(NVSS, Condon et~al. \cite{condon98}). The selection criteria and the list of
candidates have been described extensively in Paredes, Rib\'o \& Mart\'{\i}
(\cite{paredes02}), hereafter PRM02, where the search was initially limited at
low galactic latitudes $|b^{II}| \leq 5\degr$. In this paper we present
optical spectroscopy for six of the priority targets reported in PRM02, which
were also observed with the EVN and MERLIN by Rib\'o et~al. (\cite{ribo02a}).
These spectroscopic observations provide a crucial test to discern whether a
candidate is actually of stellar nature, as expected for a microquasar, or an
extragalactic object. For genuine microquasars, we expect to detect either
emission lines or absorption features of stellar photospheric origin, both
with a relatively small Doppler shift due to an eventual radial velocity of
the optical companion. Alternatively, the detection of any emission or
absorption features with a clear cosmological redshift would indicate an
extragalactic source (a quasar or other AGN).

As described herein, most of our candidates turned out to have featureless
continua; no new microquasar has been positively confirmed so far. Additional
observations are still necessary before the spectroscopic study of the full
sample is completed. Nevertheless, the negative results to date would suggest
that persistent microquasars are not a prolific nor long-lived state of
Galactic binaries.

\section{Observations} \label{observations}

The spectroscopic observations were carried out with the 4.2~m William
Herschel Telescope (WHT) at the Roque de los Muchachos observatory in Canary
Islands (Spain) in both visitor and service mode. The ISIS double-armed
spectrograph was used with different gratings in the red and blue arms, for a
resolution of 5 to 10~\AA. The technical details of this instrument are
described in the ISIS manual (Carter et~al. \cite{carter94}). The main
observations of this paper were conducted in 2002 December 2, 3, 4 and 8, with
the last night being in service mode. The observing log is listed in
Table~\ref{table:log}, which contains the source name, the observation date,
the ISIS grating used and the integration time.

\begin{table}[t]
\begin{center}
\caption[]{Log of the spectra for the high priority targets observed.}
\label{table:log}
\begin{tabular}{lccc}
\hline
\hline
1RXS Source  & Date          &   Gratings used      & Integration    \\
             & 2002          &   at the WHT         &  time (s)      \\
\hline
 J001442.2+580201   & Dec 3 & R316R, R300B   &   $3\times 2000$   \\
 J013106.4+612035   & Dec 8 & R158R, R300B   &   $1\times 1800$   \\
 J042201.0+485610   & Dec 8 &       ''       &        ''          \\
 J062148.1+174736   & Dec 2 & R316R, R300B   &   $2\times 2000$   \\
 J072259.5$-$073131 & Dec 2 &       ''       &   $2\times 2000$   \\
 J072418.3$-$071508 & Dec 8 & R158R, R300B   &   $1\times 1800$   \\
\hline
\end{tabular}
\end{center}
\end{table}

Further WHT data were also obtained on 2002 January 24, February 3 and 2003
January 21 for preliminary or verification purposes. Additional data were also
taken with the 6.5~m Multiple Mirror Telescope (MMT) and its spectrograph in
Mount Hopkins (Arizona, USA). On the MMT, we used the Blue Channel
Spectrograph with a 300 gpm grating and a $2\times 180^{\prime\prime}$ slit,
for a resolution of about 6.2~\AA. The spectra covered about 4800~\AA,
centered on 5900~\AA, and the $3072 \times 1024$-pixel ccd22 was used as a
detector. 

For each target, we obtained a series of two or three WHT spectra, with
exposure times of 1800--2000~s, with the idea of combining them during the
reduction process. The wavelength calibration was performed using arc frames
taken with Copper-Neon \& Copper-Argon lamps at the same telescope position of
the target. Arc frames were obtained before and after the science frames. A
wide range of air masses, atmospheric transparency and seing conditions was
experienced on the different nights. A similar strategy was followed at the
MMT, where we acquired the data at airmass $\sim$1.3 with seeing of about
$1.5^{\prime\prime}$, under photometric conditions. For each target, we
obtained $2 \times 1800$~s integrations; we used Helium-Neon-Argon calibration
lamps before and after each exposure. 

The data reduction was carried out using the IRAF package of NOAO including
bias subtraction, spectroscopic flat fielding, optimal extraction of the
spectra and interpolation of the wavelength solution. A few spectroscopic
standards were also observed and used to remove the spectral response and to
flux-calibrate the data. Unfortunately, the non-photometric conditions on
several nights precluded a good absolute calibration and this is most likely
accurate only within $\pm$0.3 magnitudes in some spectra.

The extracted spectra are presented in Figs.~\ref{spec1} to \ref{spec6}. The
discussion for each individual target is given in the following section. The
blue arm data of ISIS have not been included in some spectral plots due to
problems in connecting the red and blue flux calibrated data. This occurred
specially on 2002 December 2 due to bad sky transparency conditions, thus
resulting in a higher extinction at shorter wavelengths.

\section{Results on individual sources} \label{results}

Due to visibility constraints when our WHT observing time was allocated, only
the sources with Right Ascension in the 0--8~h range could be observed, i.e.,
those that were also observed with the VLA in our original paper and with the
EVN and MERLIN in Rib\'o et~al. (\cite{ribo02a}).

\subsection{1RXS~J001442.2+580201}

\begin{figure}[t]
\resizebox{\hsize}{!}{\includegraphics{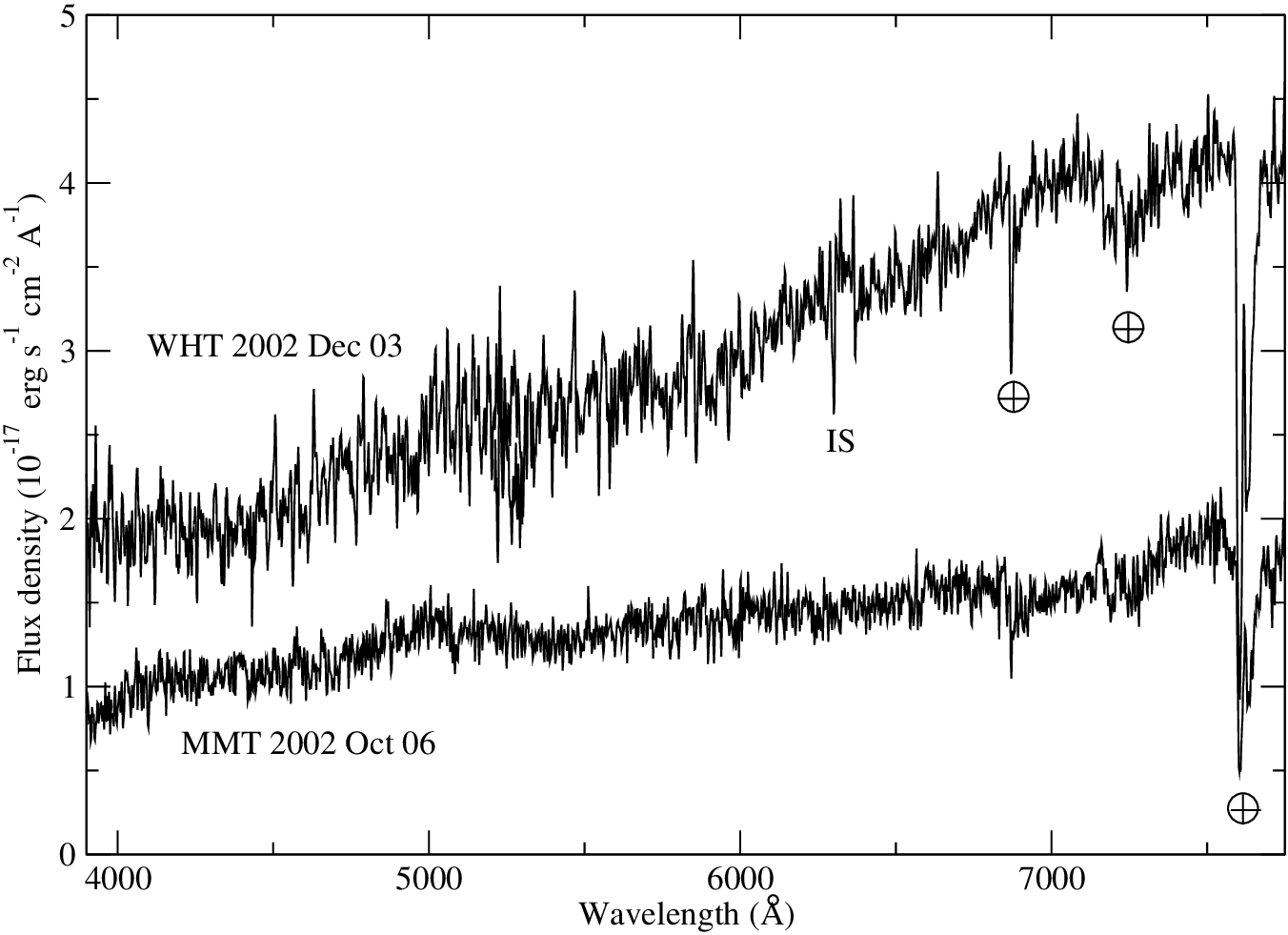}}
\caption{Optical spectra of \object{1RXS~J001442.2+580201} taken on different epochs with the WHT and the MMT. Both spectra display a simple continuum without obviously identifiable features. A possible interstellar (IS) absorption line is indicated as well atmospheric absorption bands. Such last features are more obvious in the WHT spectrum when the source was apparently brighter.}
\label{spec1}
\end{figure}

This is the weakest optical source in our sample with $I\simeq20$ mag.
\object{1RXS~J001442.2+580201} was originally considered to be a promising
candidate because of its two-sided relativistic radio jets (with $\beta>0.2$)
detected in the EVN observations by Rib\'o et~al. (\cite{ribo02a}), as well as
possible evidence of radio proper motion of the core close to the 3$\sigma$
level. However, its optical spectrum in Fig.~\ref{spec1} failed to reveal
clear emission lines or stellar photospheric features. This statement applies
to both the WHT and MMT spectra plotted in this figure. A simple continuum
with a possible interstellar and other atmospheric absorptions is detected
instead. The nature of this target remains still unknown although a galactic
stellar origin seems to be ruled out unless it is a strongly reddened star
(see discussion in Sect. \ref{redstars}). A second epoch of WHT observation,
obtained in service mode in 2003 January 21, confirmed the featureless
spectrum of this object at a comparable emission level.

Interestingly, the MMT spectrum of \object{1RXS J001442.2+580201} taken on
2002 October 6 clearly indicates that the brightness of the source was
significantly fainter by a factor of $\sim$2. The reality of this variability,
on time scales of months, could be affected by possible slit losses and other
absolute calibration uncertainties. However, if confirmed, such variations
would not be unusual for an extragalactic source.

\subsection{1RXS~J013106.4+612035}

\begin{figure}[t]
\resizebox{\hsize}{!}{\includegraphics{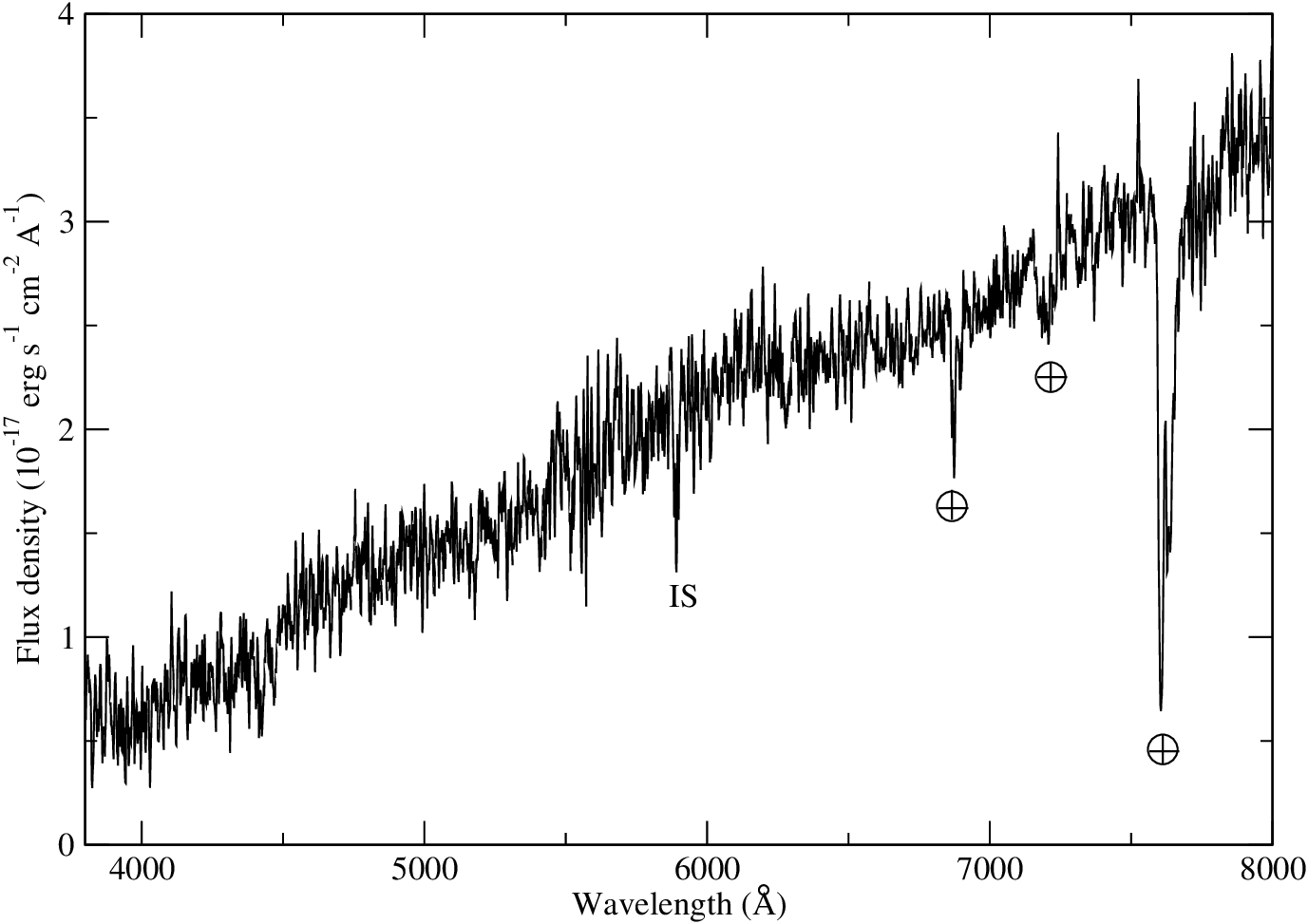}}
\caption{Optical spectrum of \object{1RXS~J013106.4+612035}.}
\label{spec2}
\end{figure}

This source was observed to exhibit a relativistic one-sided radio jet in the
EVN maps (with $\beta>0.3$). Possible proper motion of the core at the
4$\sigma$ level was also reported, although Rib\'o et~al. (\cite{ribo02a})
were very cautious about this claim. Its spectrum in Fig.~\ref{spec2} taken in
service mode displays a featureless continuum heavily absorbed at shorter
wavelengths. Only interstellar and atmospheric absorption features are
detected. An MMT spectrum is also consistent with these statements, but we do
not show it here due to likely contamination by a nearby star in the slit
already present in optical images (see Fig.~2 of PRM02). The one-sided jet and
the featureless continuum point to a possible blazar interpretation for
\object{1RXS~J013106.4+612035}. Thus, a stellar origin does not seem
appropriate for this source which is most likely extragalactic. Nevertheless,
a highly reddened star cannot be completely excluded as in the previous case.
The equivalent width (about 0.9~\AA) of the \ion{Na}{i} interstellar feature
(5890.0--5895.9~\AA), well visible for this source, does point toward a high
extinction value according to the Munari \& Zwitter (\cite{mc97}) calibration
($A_V \geq 4.5$ mag).

\subsection{1RXS~J042201.0+485610}

\begin{figure}[t]
\resizebox{\hsize}{!}{\includegraphics{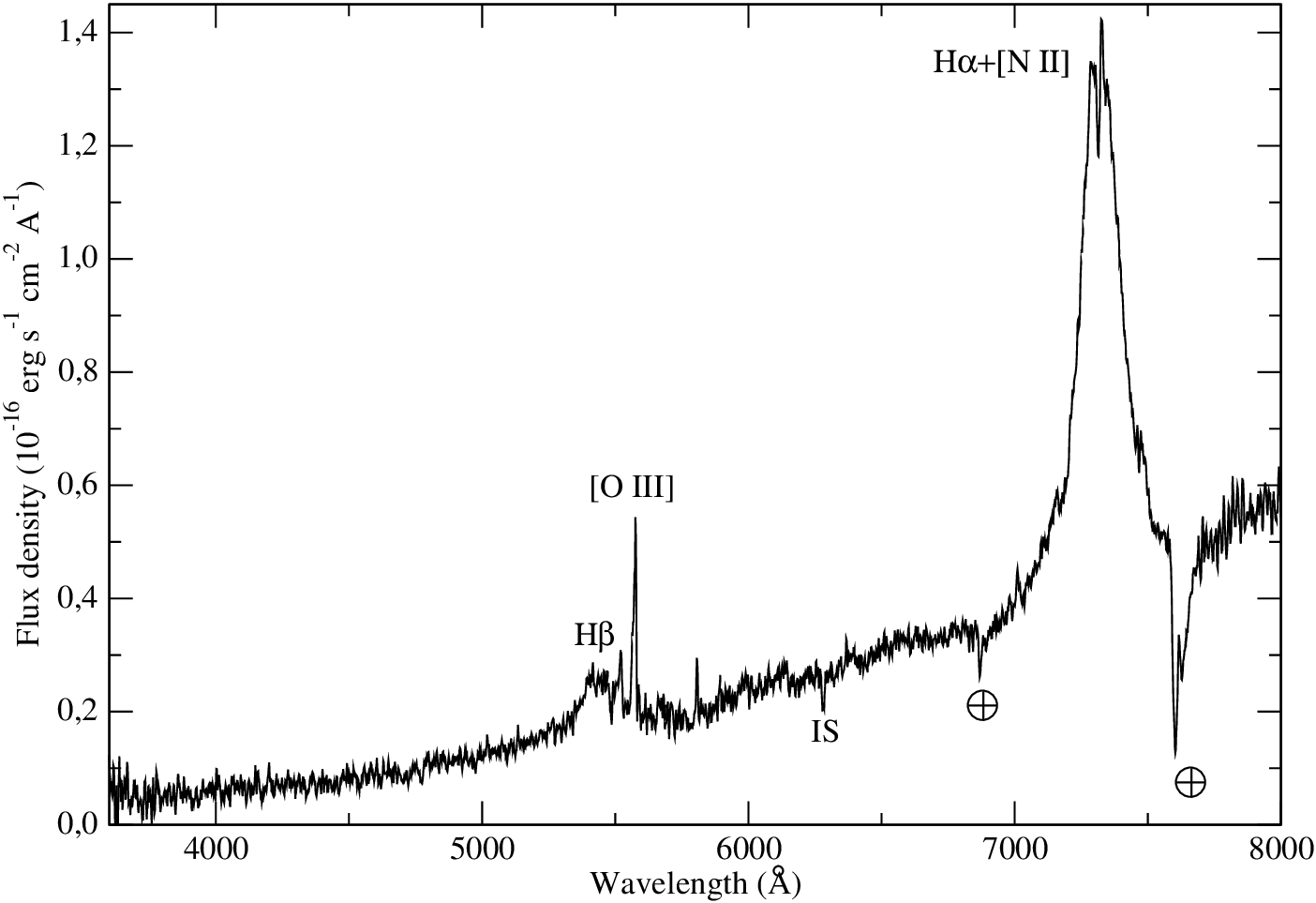}}
\caption{Optical spectrum of \object{1RXS~J042201.0+485610} with broad emission lines of hydrogen suggesting that it is a Seyfert~1 galaxy.}
\label{spec3}
\end{figure}

The spectrum of this source in Fig.~\ref{spec3} clearly shows strong emission
lines that we identify as redshifted H$\alpha$+[\ion{N}{ii}], [\ion{O}{iii}]
and H$\beta$. The H$\alpha$+[\ion{N}{ii}] blend has a broad Full Width Zero
Intensity of $\sim$20\,000~km~s$^{-1}$. Using all identified lines, the
redshift is consistently estimated to be $z=0.114\pm0.002$. An extragalactic
origin is therefore confirmed. This was already suspected in PRM02 given the 
Full Width Half Maximum (FWHM) $\sim$15\% greater than point like sources in
optical images. The broadness of the hydrogen lines suggests that this is a
likely Seyfert~1 galaxy seen through the Galactic Plane. The observed spectrum
is actually very similar to that of the Seyfert~1 galaxy and hard X-ray source
\object{GRS~1734$-$292} behind the Galactic Center (Mart\'{\i} et~al.
\cite{marti98}).

\subsection{1RXS~J062148.1+174736}

\begin{figure}[t]
\resizebox{\hsize}{!}{\includegraphics{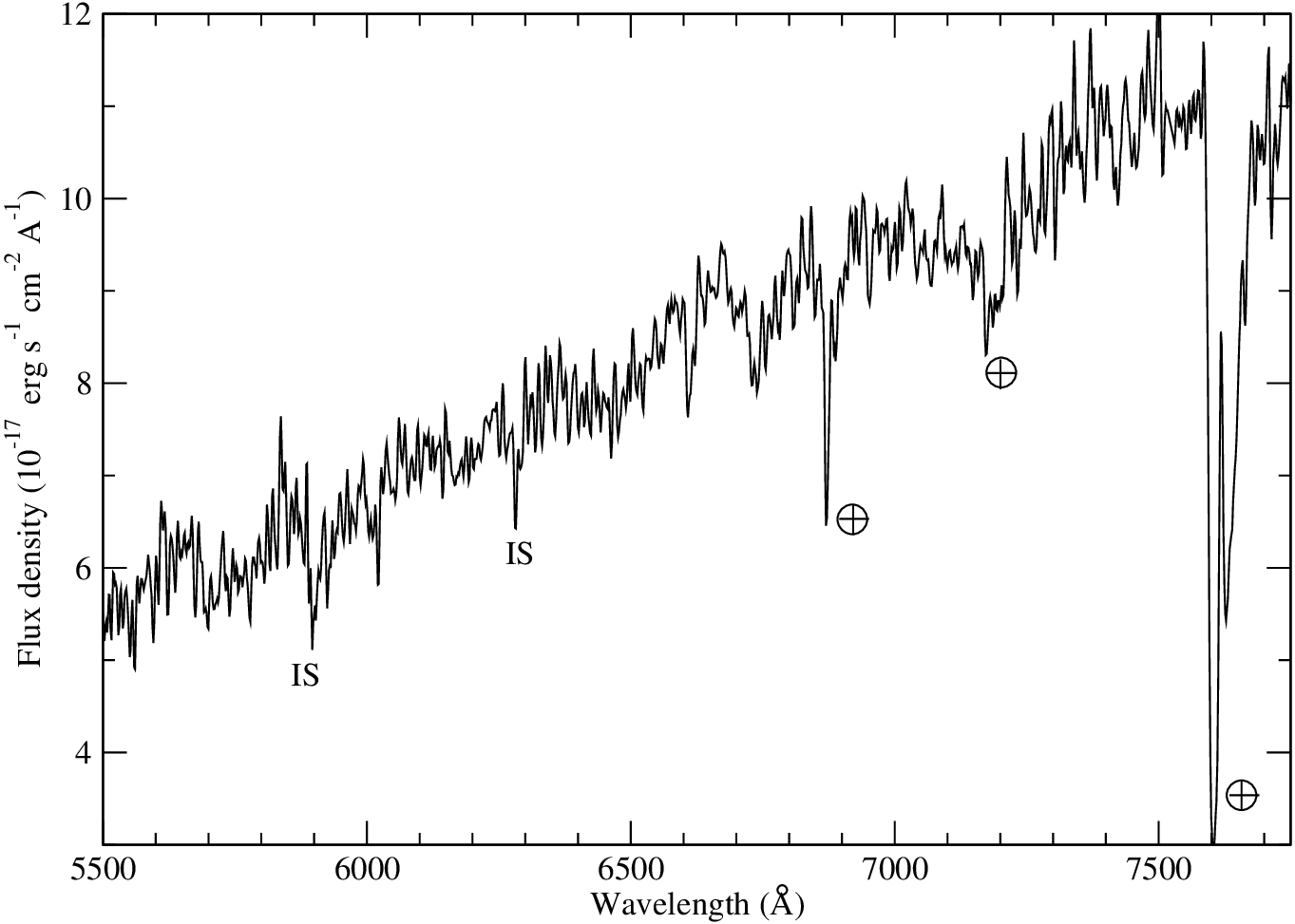}}
\caption{Optical spectrum of \object{1RXS~J062148.1+174736}. It cannot be ruled out that the apparent wiggles in this spectrum result from a sky subtraction artifact.}
\label{spec4}
\end{figure}

This source does not have any remarkable feature in the spectrum shown in
Fig.~\ref{spec4} excluding atmospheric and weak interstellar absorptions. Such
featureless continuum suggests a non-stellar origin and hence
\object{1RXS~J062148.1+174736} is likely an extragalactic object. This was
already indicated by a FWHM $\sim$30\% larger than point-like objects in the
optical images by PRM02 thus excluding possible reddening effects on a stellar
continuum.

\subsection{1RXS~J072259.5$-$073131}

\begin{figure}[t]
\resizebox{\hsize}{!}{\includegraphics{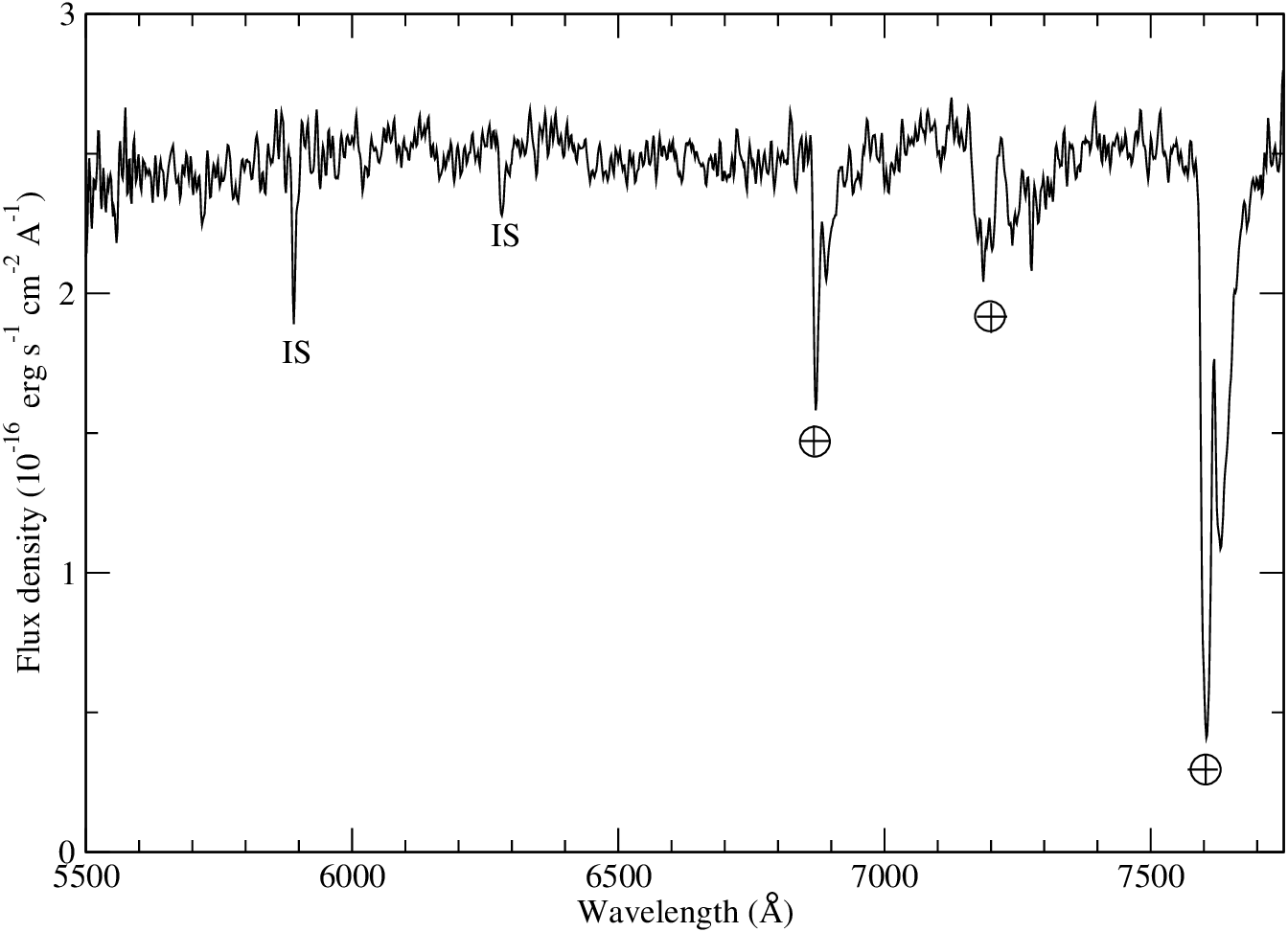}}
\caption{Optical spectrum of \object{1RXS~J072259.5$-$073131}. This object, at a galactic latitude of 3.5\degr, appears to be unusually blue. The uncertain flux calibration for the night when it was taken (Dec 2) may have affected however the slope of the continuum.}
\label{spec5}
\end{figure}

A relativistic one-sided radio jet (with $\beta>0.3$) has been detected from
this source at both arcsecond (PRM02) and sub-arcsecond scales (Rib\'o et~al.
\cite{ribo02a}). The bending of the jet in the EVN maps was reminiscent of the
ones seen in blazar sources. The featureless continuum typical of blazars that
appears in the optical spectrum of Fig.~\ref{spec5} is also in agreement with
this interpretation. \object{1RXS~J072259.5$-$073131} is therefore a likely
extragalactic source. The possiblity of a stellar reddened continuum does not
seem likely as the object appears significantly blue.

\subsection{1RXS~J072418.3$-$071508}

\begin{figure}[t]
\resizebox{\hsize}{!}{\includegraphics{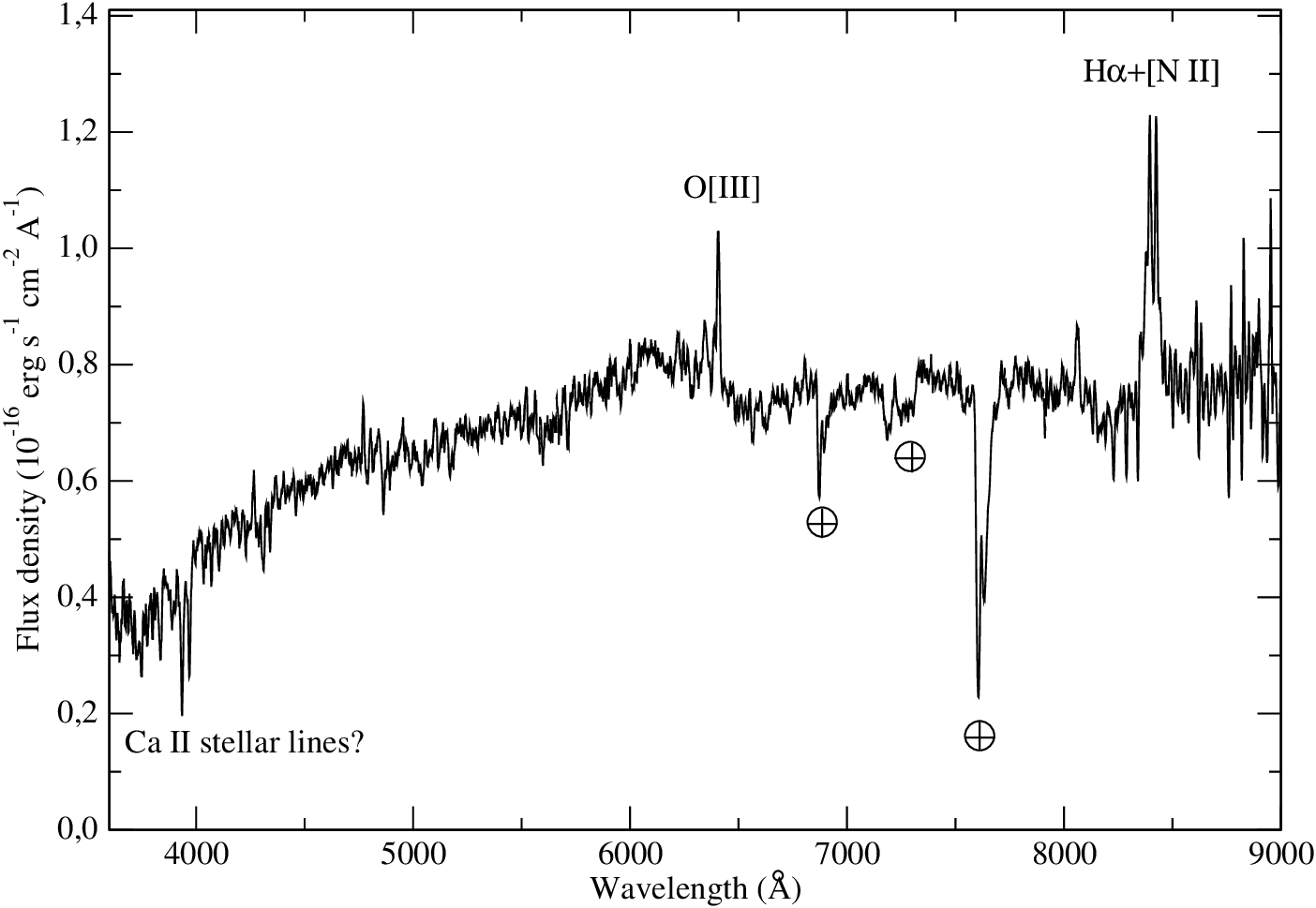}}
\caption{Optical spectrum of the flat spectrum radio quasar \object{1RXS~J072418.3$-$071508}. The \ion{Ca}{ii} absorption lines are likely to be due to a superposed late type star.}
\label{spec6}
\end{figure}

This source was considered a candidate when we first performed VLA and optical
observations at the early stages of our research. However, around this epoch
the source was classified as a flat spectrum radio quasar by Perlman et~al.
(\cite{perlman98}) with a redshift of $z=0.270$. Our spectrum in
Fig.~\ref{spec6} shows the presence of strong, redshifted emission lines of
H$\alpha$+[\ion{N}{ii}] and [\ion{O}{iii}] for this relativistic one-sided
radio jet source with $\beta>0.5$ (Rib\'o et~al. \cite{ribo02a}). We derive a
redshift estimate of $z=0.280\pm0.001$, not very different from the previous
finding. The small discrepancy is probably due to the different
signal-to-noise ratio between our spectrum and that of Perlman et~al.
(\cite{perlman98}). H$\alpha$+[\ion{N}{ii}] were noticeably stronger at the
epoch of our observations. The presence of \ion{Ca}{ii} absorption lines close
to their rest wavelength in Fig.~\ref{spec6} could be explained if a late type
star is almost superposed on the quasar line of sight. The presence of this
star is inferred from the double appearance of this source in CCD optical
images (see Fig.~2 of PRM02).

\section{Discussion} \label{discussion}

The selection of microquasar candidates by PRM02 was mostly sensitive to
persistent systems, especially in high mass X-ray binaries (HMXBs). Indeed,
the selection algorithm recovered well known microquasars such as
\object{SS~433}, \object{Cygnus~X-3} and \object{LS~5039}, as well as the Be
HMXB \object{LS~I~+61~303} (also a source of relativistic jets as observed
with the EVN and MERLIN; Massi et~al. \cite{massi01}; Massi et~al.
\cite{massi03}). The selection algorithm did not recover any low mass X-ray
binary (LMXB). This can be understood because LMXBs tend to be transient
sources at both X-ray and radio wavelengths, hence not being generally present
in the NVSS nor the RBSC.

Unfortunately, however, none of the six priority candidates in PRM02 appears
to be clearly of galactic stellar nature. One of them is a Seyfert galaxy and
another a significantly redshifted quasar. The other four exhibit a
featureless continuum with only interestellar absorption lines superimposed on
it, reminiscent of BL~Lac objects or blazars. Such extragalactic
interpretation seems more likely for \object{1RXS~J062148.1+174736} and
\object{1RXS~J072259.5$-$073131}, whose spectra have a relatively good 
signal-to-noise ratios (SNR) of about 30 and 50, respectively. Indeed, at this
SNR, a stellar spectrum should normally be recognizable and even a stellar
spectral type derived. This is clearly not the case for these two objects.

Nevertheless, it could still be possible that the remaining two featureless
sources are highly reddened stars seen through the disc of the Galaxy, as
already advanced in the previous section. This statement applies to
\object{1RXS~J001442.2+580201} and \object{1RXS~J013106.4+612035} whose
observed spectra are of poorer quality.

\subsection{Highly reddened stars?} \label{redstars}

\begin{figure}[t]
\resizebox{\hsize}{!}{\includegraphics{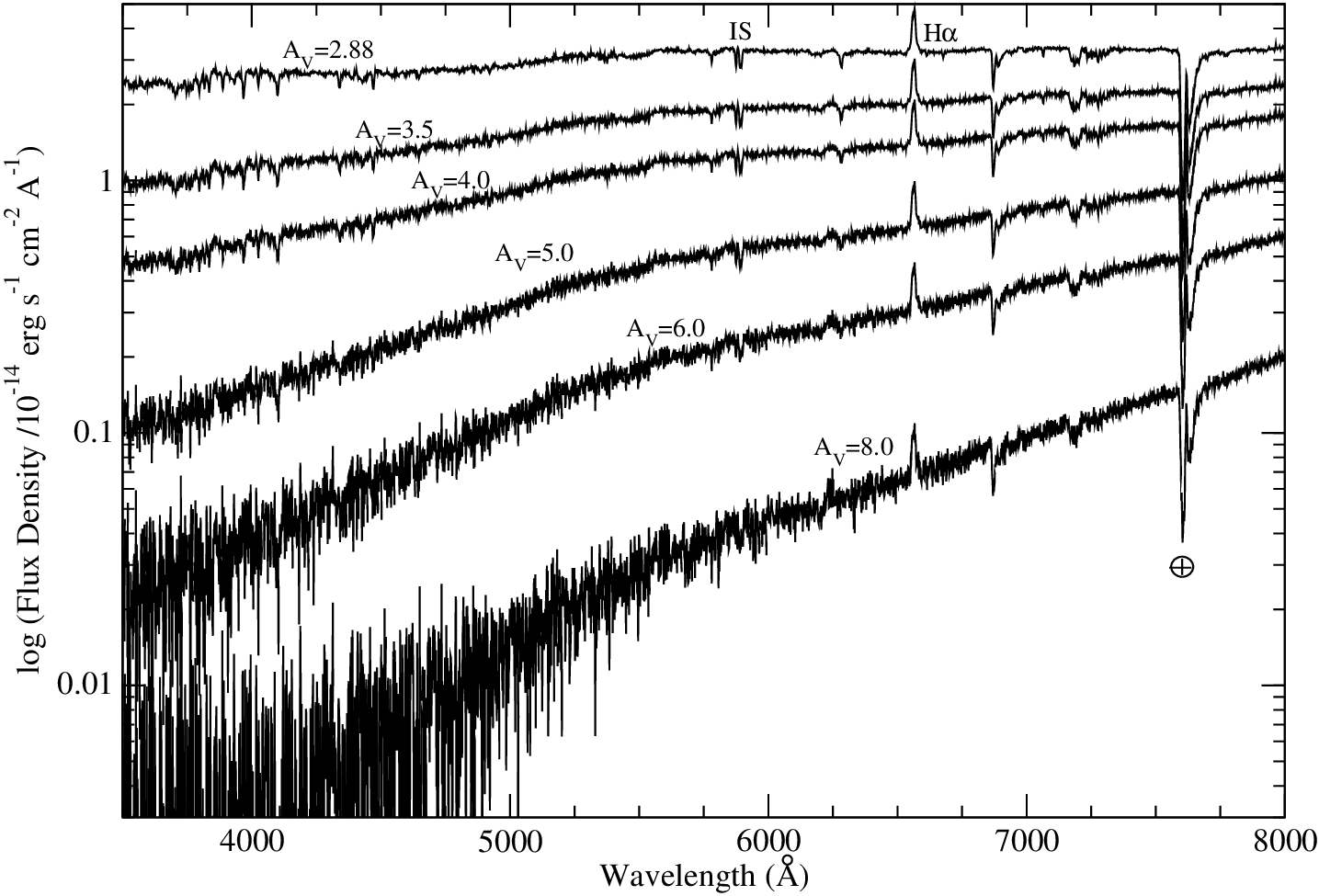}}
\caption{Optical spectra of the microquasar \object{LS~I~+61~303} as would be observed with increasing values of the extinction at optical wavelengths assuming constant exposure time. The 120~s spectrum at the top of the plot was originally taken with the WHT and ISIS on 2002 December 2 ($A_{\rm V} = 2.88$). Logarithmic vertical scale is used to better illustrate how the photospheric lines of the stellar companion become progressively difficult to detect within noise as the extinction and reddening increase. Only the strong H$\alpha$ emission line of the system would eventually remain as a conspicuous feature.}
\label{red}
\end{figure}

In order to consider the possibility just quoted above, we have carried out
the exercise of reddening the optical spectrum of a HMXB microquasar such as
\object{LS~I~+61~303} in a progressive way. The intrinsic reddening towards
this system is assumed to be $E(B-V)=0.93$ mag, equivalent to a visual
extinction of $A_V=2.88$ mag (Hutchings \& Crampton \cite{hc81}). The IRAF
task {\sc deredden} was used for this purpose. The original spectrum of
\object{LS~I~+61~303} was taken using the WHT with the same instrumental setup
as for our candidate sources. In Fig.~\ref{red}, we display the different
spectra of this microquasar as seen with increasing values of extinction.
Noise was artificially added with the IRAF task {\sc mknoise} to approximately
simulate the decaying SNR expected while the flux density decreases
progressively with reddening. Exposure time is assumed to remain constant. It
is clear from this plot that the many photospheric lines of the early type
star become less and less evident as $A_V$ increases, specially in the blue
part of the spectrum. The most reddened spectra in Fig.~\ref{red} have a
SNR$\sim$10 in this region, i.e., comparable to that of
\object{1RXS~J001442.2+580201} and \object{1RXS~J013106.4+612035} in our real
observed spectra. Therefore, we cannot strictly rule out the possibility of
highly reddened stars in addition to the BL Lac interpretation for these two
featureless sources. Addtional spectroscopic observations, with higher SNR (or
at infrared wavelenghts), would be necessary to distinguish between the two
interpretations.

\subsection{The extragalactic alternative} \label{extragalactic}

If the ambiguous sources are not highly reddened stars, then all our
candidates would turn out to be extragalactic sources and no new microquasar
would be present among them. Although some secondary candidates remain to be
explored, a non-detection of new microquasars is nevertheless noteworthy.

Assuming that all our sources are extragalactic, it is possible to estimate an
upper limit for the density of new persistent microquasar systems in the
Galaxy. In order to do so, we need to assess the galactic regions accessible
to our search as follows. The NVSS and the RBSC come from surveys obtained in
different epochs separated by few years. Hence, a microquasar is most likely
to be present in both of them in case it is a persistent source over the
years. The NVSS was conducted at the 20~cm wavelength with the VLA and its
brightness limit is such that sources with flux density of $\sim$2.5~mJy are
the weakest reliable detections.

The RBSC was compiled using PSPC data from the ROSAT satellite, in the energy
range 0.1--2.4~keV, and its brightness limit is about 0.1 count~s$^{-1}$. This
limit corresponds to an unabsorbed energy flux that depends on the source
spectrum and the amount of hydrogen column density $N_{\rm H}$ towards it. 
The column density can be estimated using the Predehl \& Schmitt
(\cite{predehl95}) relationship with the extinction at visual wavelengths
$A_V$. We will further assume a crude but plausible value of 1.9
mag~kpc$^{-1}$ for the total absorption of star light near the galactic plane
(Allen \cite{allen73}), that is:
\begin{equation}
N_{\rm H} = 1.79 \times 10^{21}~{\rm cm}^{-2} A_V = 3.40 \times 10^{21} D~{\rm cm}^{-2}
\label{extin}
\end{equation}
where $D$ is the distance to the source in kpc. 

A persistent microquasar like \object{Cygnus~X-3} typically has a 20~cm flux
density of $\sim$100~mJy at about 10~kpc from the Sun. Therefore, it would be
detected in the NVSS up to a distance $\leq 63$~kpc since the Galaxy is almost
transparent at cm wavelengths. The situation is more restrictive in the ROSAT
energy range. Here, the plausible persistent X-ray luminosity of a
\object{Cygnus~X-3} system would be $L_{\rm X} \sim 10^{37}$ erg~s$^{-1}$ and
we will further assume a power law index $\Gamma=+2$. For such object to
appear detected in the RBSC above the 0.1 count~s$^{-1}$ level, its distance
needs to be $\leq 15$~kpc. This number has been derived using the web based
tool PIMMSv3.3d originally developed by Mukai (\cite{mukai93}) for flux and
count conversion between different X-ray observatories. To do so, from
Eq.~\ref{extin}, the plausible hydrogen column density at 15~kpc is crudely
estimated as $N_{\rm H} = 5.1 \times 10^{22}$~cm$^{-2}$. Then, the unabsorbed
energy flux corresponding to 0.1 count~s$^{-1}$ according to PIMMSv3.3d is
$4.2 \times 10^{-10}$ erg~cm$^{-2}$ s$^{-1}$, which is consistent with the
assumed X-ray luminosity when multiplied by $4\pi D^2$.

By imposing a detection at both radio and X-rays, the microquasar search using
the PRM02 criteria is sensitive to \object{Cygnus~X-3} like systems within the
second (smaller) distance limit of $D=15$~kpc. This includes a substantial
volume of the Galaxy, which can be estimated as:
\begin{equation}
V = {4\pi \over 3} \eta D^3 \sin{|b^{II}|_{\rm max}}, \label{volume}
\end{equation}
where $\eta$ is the fraction of the Galactic Plane sampled in our search and
$|b^{II}|_{\rm max}=5\degr$ its limit in galactic latitude. The fraction
$\eta$ is set mainly by the declination limit ($\delta \geq -40\degr$) in
PRM02, which only allowed to search within galactic logitudes
$-$13.3\degr$\leq l^{II} \leq$259.2\degr. This represents 75.7\% of the
Galactic Plane. However, our search close to the Galactic Center (roughly
$-1\degr \leq l^{II} \leq 1\degr$) is severely affected by source confusion
specially in the NVSS. This reduces the sampled fraction by 0.55\% only, but
it is very likely that many interesting sources have been missed in these
central regions of the Galaxy. Excluding this caution, we have $\eta \simeq
0.75$ and consequently $V \simeq 920$~kpc$^3$. The fact that not even one
\object{Cygnus~X-3}-like system has been apparently discovered suggests that
bright (and persistent) microquasars are probably not very abundant in the
galactic disc. The corresponding upper limit for new \object{Cygnus~X-3}
systems ($\leq 1/V$) is about $\la 1.1 \times 10^{-12}$ pc$^{-3}$.

There is, of course, the alternative of a persistent but fainter microquasar
population which could have been missed by our search. In this case, how far
away could we detect such systems? Let us assume that these objects are
comparable to \object{LS~5039}, which is a relatively faint system at both
X-ray ($L_{\rm X} \sim 5 \times 10^{34}$ erg~s$^{-1}$) and radio ($\sim$20~mJy
at 20~cm) wavelengths at $2.9\pm0.3$~kpc from the Sun (Paredes et~al.
\cite{paredes00}, Rib\'o et~al. \cite{ribo99}, \cite{ribo02b}).

Under such assumptions, and proceeding just as above, such faint and
persistent microquasars would be detected in the NVSS for distances $\leq
8.5$~kpc and in the RBSC for distances $\leq 4$~kpc. A simultaneous detection,
in both surveys, would be thus possible only for relatively nearby systems
within $D=4$~kpc from the Sun. Again, no new system of this kind has been
apparently discovered in our search. If they exist, their population in the
vicinity of the Sun does not seem to be very numerous as well, at least close
to the galactic plane. Using again Eq.~\ref{volume} we have a sampled volume
of $V \simeq 18$~kpc$^3$ in this case, with the corresponding density upper
limit for new \object{LS~5039}-like systems being $\la 5.6 \times
10^{-11}$~pc$^{-3}$. 

However, there may be several faint and persistent microquasars at larger
distances in the Galaxy where our search is not sensitive to them. On the
other hand, it may also well happen that nearby systems are naturally located
at galactic latitudes higher than what we have been searching for and a more
extended search is then necessary. Similarly, higher galactic latitudes for
\object{LS~5039}-like systems appear also conceivable as they may be runaway
microquasars escaping from the galactic disc (Rib\'o et~al. \cite{ribo02b}).

In this context, we have just started to expand our cross-identification
studies to $5^{\circ} \leq |b^{II}| \leq 10\degr$. This is not a simple task
considering the increasing amount of sky to be covered, which provides a large
number of sources to be explored. Results from this expanded search will be
reported in other papers. Future searches would also benefit from sensitive
surveys in gamma and hard X-rays, such as the current INTEGRAL Galactic Plane
Survey and the planned EXIST mission.

\section{Conclusions} \label{conclusions}

\begin{enumerate}

\item Spectroscopic observations have been reported for six of the high
priority microquasar candidates reported in PRM02 within galactic latitude
$|b^{II}| \leq 5\degr$. Two of them are clearly extragalactic objects while
the remaining four display a featureless continuum in their spectra.

\item The four candidates with featureless spectra are reminiscent of BL Lac
or blazars and would be therefore extragalactic sources as well. However, the
possibility that two of them (\object{1RXS~J001442.2+580201} and
\object{1RXS~J013106.4+612035}) are highly reddened stellar systems cannot be
completely excluded based on the present data.

\item Assuming that none of our candidates is galactic, it appears that the
population of new and persistent microquasars is not very numerous in the
Galaxy. The corresponding density of new \object{Cygnus~X-3} and
\object{LS~3039}-like system is constrained to be $\la 1.1 \times 10^{-12}$
pc$^{-3}$ and $\la 5.6 \times 10^{-11}$~pc$^{-3}$, respectively. 

\end{enumerate}

\begin{acknowledgements}
The William Herschel Telescope is operated on the island of La Palma by the
Isaac Newton Group in the Spanish Observatorio del Roque de los Muchachos of
the Instituto de Astrof\'{\i}sica de Canarias.
J.~M., J.~M.~P. and M.~R. acknowledge partial support by DGI of the
Ministerio de Ciencia y Tecnolog\'{\i}a (Spain) under grant AYA2001-3092.
J.~M. has been also aided in this work by an Henri Chr\'etien International
Research Grant administered by the American Astronomical Society.
J.~M. is also supported by the Junta de Andaluc\'{\i}a (Spain) under project
FQM322.
M.~R. acknowledges support from a Marie Curie individual fellowship under
contract No.~HPMF-CT-2002-02053.
J.~S.~B is partially supported by a discrentionary grant from the
Harvard-Smithsonian Center for Astrophysics and also thanks 
the University of Barcelona for a generous travel grant.
We thank as well A.~Soderberg for kindly taking some imaging data at the
Palomar 60-inch and J.~E. McClintock for useful comments to the manuscript.
We finally wish to acknowledge X. Barcons (CSIC-UC), and the service observers of the Isaac Newton Group, for kindly obtaining some of the spectra for this
work.
Finally, P.-A. Duc (CEA/Saclay) and M. S\'anchez G\'omez (UJA) are also acknowledged for useful comments in the data analysis.
\end{acknowledgements}

\end{document}